# Anomalous Hall and Nernst effects in the Two-Dimensional ferromagnetic metal FePd$_2$Te$_2$

Yazhou Li[1,#], Jiaxing Liao[1,#], Jiajun Ma[1], Yuwei Zhang[1], Tao Li[1], Jialu Wang[1], Hangdong Wang[1], Hanjie Guo[2], Jianhui Dai[1], Yuke Li[1,*]

1. School of Physics and Hangzhou Key Laboratory of Quantum Matters, Hangzhou Normal University, 311121, China
2. *Neutron Science Platform, Songshan Lake Materials Laboratory, Dongguan, Guangdong 523808, China*

## Abstract

The transverse thermoelectric effect enables simpler, more flexible thermoelectric devices by generating electricity perpendicular to heat flow, offering promising solutions for waste heat recovery and solid-state cooling applications. Here, we report a striking observation of zero-field anomalous Hall effect (AHE) and anomalous Nernst effect (ANE) below $T_C$ in the two-dimensional metallic magnet FePd$_2$Te$_2$. The anomalous Nernst signal $S_{yx}^A$ peaks a maximum value of 0.15 $\mu V/K$ at 100 K, much larger than that of conventional FM materials. Remarkably, the derived ratio $|\alpha_{ij}^A/\sigma_{ij}^A|$ in FePd$_2$Te$_2$ approaches the fundamental limit of $k_B/e = 86\ \mu V/K$. Our findings suggest a dominant Berry curvature contribution to the ANE. The observed giant zero-field anomalous Nernst response in 2D FePd$_2$Te$_2$ not only advances fundamental understanding of transverse thermoelectricity in layered magnets, but also provides this material as a promising candidate for practical thermoelectric spintronic applications.

# Introduction

Two-dimensional (2D) magnetic materials have become a central research focus in condensed matter physics due to their unique quantum phenomena and transformative potential for spintronic applications[1-6]. In particular, for 2D magnetic metals, the interplay of itinerant electrons and long-range magnetic ordering creates unique opportunities to explore emergent quantum phenomena and develop novel spintronic devices[5, 7-9]. Unlike their insulating or semiconducting compounds[10], including well-studies examples: $Cr_2Ge_2Te_6$[11], $FePS_3$[12], CrSBr[13] and $CrI_3$[14], these metallic systems exhibit intrinsic spin-charge coupling through conduction electron-mediated magnetic interactions[5, 7-9], offering distinct advantages for both fundamental research and device applications. In this context, $Fe_3GeTe_2$ stands out as a rare vdW ferromagnetic(FM) metal whose rich topological electronic structure has unveiled a spectrum of novel quantum phenomena[15-18], including the Anomalous Quantum Hall effect (AQHE) [19], Anomalous Nernst effect (ANE) [10, 20], and Skyrmions[21, 22].

Recently, the discovery of the vdW magnetic metal $FePd_2Te_2$ has established a promising platform for investigating 2D-quantum materials[23]. This system possesses several distinctive characteristics: (1) remarkable air stability and mechanical exfoliability down to few-layer thicknesses; (2) strong in-plane uniaxial magnetic anisotropy with a FM transition temperature of $T_C$ = 183 K, making it suitable for integration into functional vdW heterostructures; and (3) unique spin textures with perpendicular domain boundaries arising from twinning crystal effects, which may host emergent quantum phenomena including anomalous Hall and Nernst effects. This unique combination of properties positions $FePd_2Te_2$ as an ideal system for both fundamental investigations of low-dimensional magnetism and the development of advanced vdW-based quantum devices.

Here, we performed a detailed investigation of its magnetic properties and anomalous transverse thermoelectric effects in the 2D ferromagnet metal $FePd_2Te_2$. It shows a remarkable zero-field AHE and ANE below $T_C$, exhibiting a significant hysteresis loop. The anomalous zero-field Nernst signal $S_{yx}^A$ attains a maximum value of 0.15 $\mu V/K$ at 100 K, while the derived ratio $|\alpha_{ij}^A/\sigma_{ij}^A|$ is close to the fundamental limit of $k_B/e = 86 \mu V/K$. These findings suggest dominant Berry curvature contributions to the anomalous transverse transport coefficients. This work establishes $FePd_2Te_2$ as an exemplary platform for investigating anomalous transport properties in

2D ferromagnets and a promising materials system for developing potential thermoelectric spintronic devices.

## Experimental methods

$FePd_2Te_2$ single crystals were grown via a solid-state reaction method. We mixed and ground powders of Fe, Pd, and Te in a molar ratio of 1:2:2, and then sealed them in a quartz tube under vacuum conditions. The sealed quartz tube was placed in a box furnace and heated at 800°C for 2 days. After that, we slowly decrease the temperature to 600°C at a cooling rate of 2°C/h. Finally, the furnace was turned off and naturally cooled to room temperature. The obtained $FePd_2Te_2$ single crystals show typical dimensions of 3 × 4 × 0.5 mm³.

The chemical composition of crystals was determined by energy dispersive X-ray spectroscopy (EDS) measurements, revealing a Fe: Pd: Te ratio of approximately 1:2:2. The flat plane of the single crystals was determined by X-ray diffraction (XRD) as $(10\bar{1})$. Magnetic properties were characterized using a SQUID system. The thermoelectric transport measurements were performed using a 14T-liquid-helium-free superconducting magnet system manufactured by Cryogenic Ltd. (UK). Thermopower and Nernst effect were simultaneously measured with a one-heater–two-thermometer technique in a 14T-cryogenic refrigerator with a high-vacuum environment.

The Hall and Nernst measurements can be influenced by slight misalignment of voltage contacts, which may introduce longitudinal signal contributions. Therefore, the signals are antisymmetrized with respect to the magnetic fields to eliminate this misalignment.

# Results and discussion

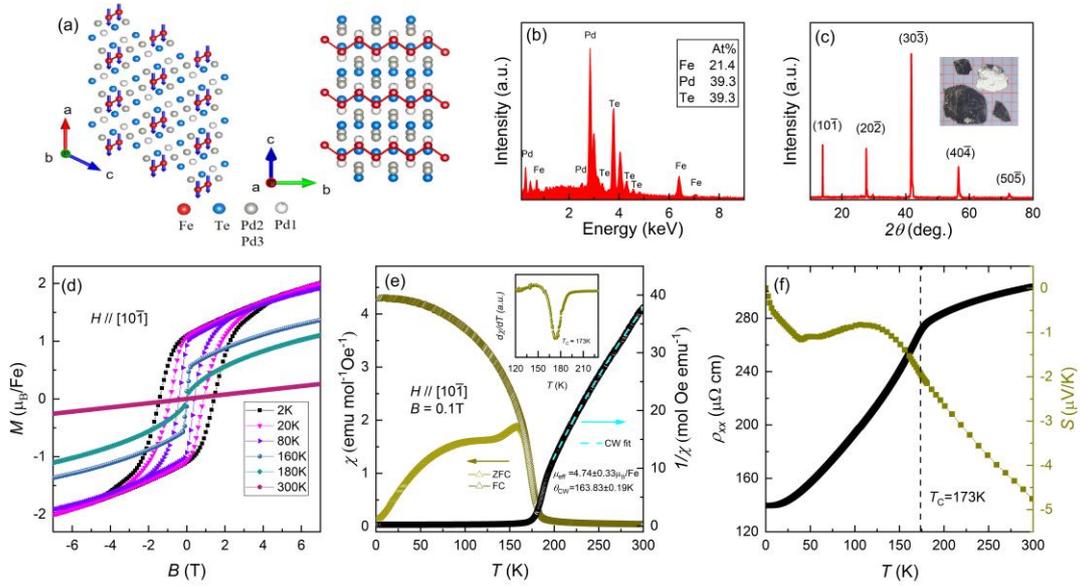

Figure 1 Characterization analysis of FePd$_2$Te$_2$ single crystal: (a) Crystal structure and spin arrangement. (b) EDX spectrum. (c) X-ray diffraction spectrum. The inset shows a photograph of a single crystal. (d) Magnetization curves vs magnetic fields at different temperatures. (e) Temperature dependence of the magnetization (left axis) and the inverse of the magnetization (right axis) of FePd$_2$Te$_2$ at a magnetic field of $B = 0.1$ T parallel to the $(10\bar{1})$ plane. The dashed line represents the Curie-Weiss fitting results with temperature range from 200 to 300 K. Inset: Derivative of magnetization with respect to temperature; (f) Resistivity (left plane) and thermopower (right plane) as functions of temperature.

The van der Waals ferromagnetic metal FePd$_2$Te$_2$ crystallizes in a monoclinic crystal structure with a layered two-dimensional configuration, exhibiting a strong cleavage tendency along the $(10\bar{1})$ plane[23]. This is further confirmed by a series of $(l0\bar{l})$ peaks observed in the XRD pattern (Figure 1b). The Energy Dispersive X-ray (EDX) spectrum measurement in Figure 1a shows an Fe:Pd:Te atom ratio of 21.4:39.4:39.4, matching the expected stoichiometry of FePd$_2$Te$_2$. The temperature dependence of the in-plane magnetization under $B = 0.1$ T rapidly increases below the Curie temperature $T_C$, exhibiting a typical ferromagnetic (FM) behavior. The $T_C$ of nearly 173 K, as shown in the inset of Figure 1c, is much higher than the previous results[23]. The difference could be ascribed to the different iron contents in FePd$_2$Te$_2$, similar to the case of Fe$_3$GeTe$_2$[20, 24, 25]. The $\chi$ at high temperature from 200 to 300 K follows the modified Curie-Weiss law[25]: $\chi = C/(T-\theta) + \chi_0$, where $C$ is the Curie constant, $\theta$ is the Curie-Weiss temperature, and $\chi_0$ is a temperature-independent term mostly from the Pauli paramagnetism. The obtained $\theta$ and effect moment are close to 163.8 K and $4.74\mu_B$/Fe, respectively, suggesting the presence of ferromagnetic coupling between $Fe^{2+}$ ions. The resistivity as a function of temperature in Figure 1d shows a linear decrease from room temperature, but quickly drops below $T_C$, due to the decreasing impurity scattering originating from the electron spin alignment. The negative Seebeck coefficient $S_{xx}$ in Figure 1d indicates the dominant electron-type carriers. Below TC, the Seebeck coefficient deviates from linear temperature-dependent, showing a local minimum in absolute value near 115 K, resembling that observed in ferromagnetic Weyl semimetal $Co_3Sn_2S_2$[26]. The observed local minimum in the Seebeck coefficient likely stems from either the suppression of spin-disorder scattering due to ferromagnetic ordering or enhanced phonon-drag effects mediated by magnon-phonon coupling.

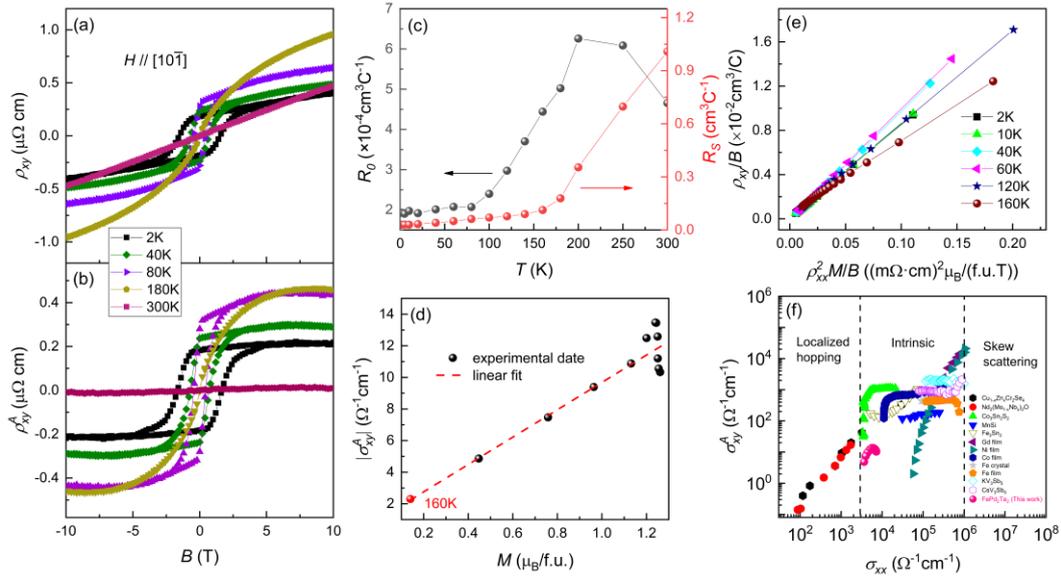

Figure 2 (a) The magnetic field dependence of Hall resistivity at various temperatures for $H \parallel [10\bar{1}]$ direction. (b) Anomalous Hall effect. (c) Temperature dependence of the normal Hall coefficient $R_0$ (left axis) and the anomalous Hall coefficient $R_S$ (right axis). (d) The anomalous Hall conductivity $|\sigma_{xy}^A|$ vs. the magnetization $M$ below $T_C$. (e) The relationship between $\rho_{xy}/B$ and $\rho_{xx}^2 M/B$; (f) A unified model of the anomalous Hall conductivity $\sigma_{xy}^A$ as a function of $\sigma_{xx}$ for different magnetic materials. $\sigma_{xy}^A$ is extracted below $T_C$ in $FePd_2Te_2$.

Hall resistivity $\rho_{xy}$ of $FePd_2Te_2$ as a function of magnetic field in Figure 2a exhibits a significant resistivity loop at low fields, implying an anomalous Hall component $\rho_{xy}^A$, consistent with the behavior observed in the magnetization curve (Figure 2b). At high fields, Hall resistivity tends to linearly increase, denoting the normal Hall effect $\rho_{xy}^N$. In this case, the Hall resistivity can be described by the following relation:

$$\rho_{xy} = \rho_{xy}^N + \rho_{xy}^A = R_0 B + 4\pi R_S M \qquad (1)$$

$R_0$ and $R_S$ represent the normal and anomalous Hall coefficients. Through subtracting the linear term (the normal Hall effect) at high fields, the anomalous $\rho_{xy}^A$ can be derived in Figure 2b. The anomalous Hall resistivity reaches ~0.45 $\mu\Omega.cm$, comparable to the value in the topological antiferromagnet FeGe[27]. The obtained $R_0$ and $R_S$ rapidly decrease with temperature decreasing from 300 K, but suddenly turn to remain less changed below 100 K and 180 K, respectively, as described in Figure 2c. The calculated $R_S$ is about three orders of magnitude larger than $R_0$, indicating the dominant AHE in FePd$_2$Te$_2$.

Within the unified theoretical framework[28], the AHE can originate from either extrinsic mechanisms (magnetic impurity scattering) or intrinsic mechanisms (Berry curvature effects). The extrinsic contribution comprises both side-jump and skew scattering processes, while the intrinsic AHE is described by $\sigma_{xy}^A = -\frac{e^2}{\hbar}\int \frac{d^3k}{(2\pi)^3} f(k)\Omega_{xy(k)}$, where $f(k)$ represents the Fermi-Dirac distribution and $\Omega_{xy(k)}$ is the Berry curvature along the xy-plane[29]. Here, we investigate the origin of the AHE in FePd$_2$Te$_2$ through examining the dependence of the anomalous Hall conductivity $|\sigma_{xy}^A|$ on magnetization $M$ in Figure 2d. At 60 K < $T$ < $T_C$, the $|\sigma_{xy}^A|$ exhibits a linear relationship with $M$, consistent with the intrinsic Kubo-Luttinger (KL) mechanism[30], as previously reported in La$_{0.7}$Sr$_{0.3}$MnO and Fe$_3$GeTe$_2$[25, 31, 32]. However, below 60 K, the experimental data deviate from the linear trend, possibly due to magnetic impurity scattering[25]. We thus plot the $\rho_{xy}/B$ vs. $\rho_{xx}^2 M/B$ for different temperatures in Figure 2e. The observed linear relationship below $T_C$ supports the intrinsic KL mechanism as the dominant contribution to the AHE. The intrinsic contribution can also be identified by plotting $|\sigma_{xy}^A|$ as a function of the longitudinal conductivity $\sigma_{xx}$, as shown in Figure 2f. The Hall resistivity of FePd$_2$Te$_2$ clearly resides within the intrinsic regime, providing another evidence that Berry curvature dominates the AHE in this system.

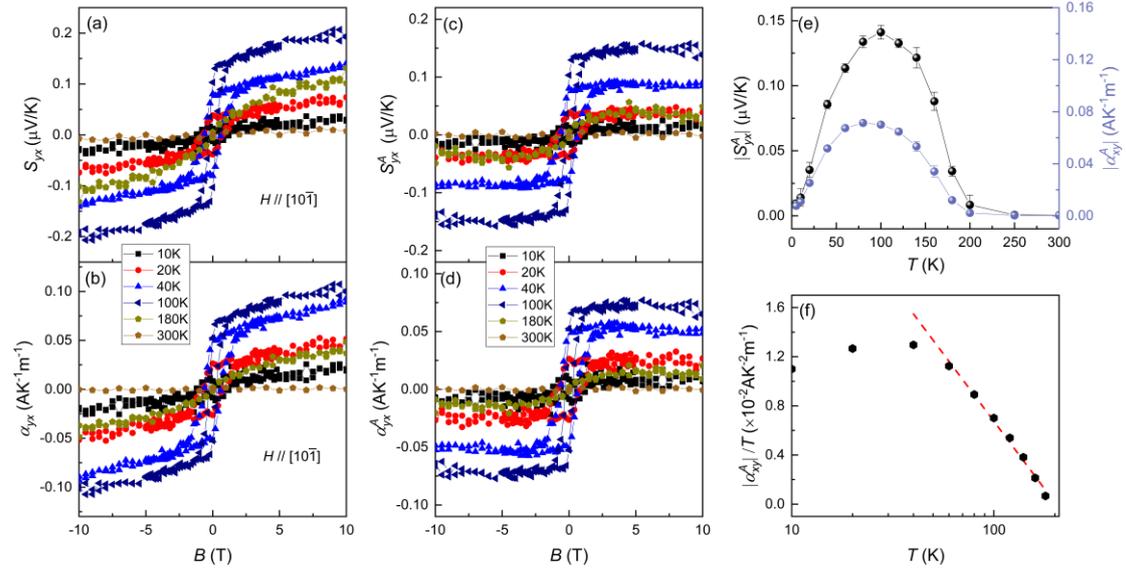

Figure 3 (a) The Nernst coefficient $S_{yx}$ and (b) the transverse thermoelectric conductivity $\alpha_{yx}$ as function of magnetic field at different temperatures as $H \parallel [10\bar{1}]$. (c) Temperature dependence of the anomalous Nernst coefficient $S_{yx}^A$ (left axis) and anomalous thermoelectric conductivity $\alpha_{xy}^A$ (right axis); (d) $|\alpha_{xy}^A|/T$ as a function of temperature.

The Nernst effect provides a powerful probe for investigating intrinsic transport properties. We thus perform the Nerst effect measurements in FePd$_2$Te$_2$, as shown in Figure 3a. Below $T_C$, the Nernst signal S$_{yx}$ shows a clear loop at low fields and tends to increase linearly at high fields. With the increasing temperature to 100 K, the $S_{yx}$ increases to its maximum value of ~ 0.2 $\mu V/K$. Like the case of Hall effect, in a FM system, the total Nernst signal is composed of the normal ($S_{yx}^N$) and anomalous Nernst effect ($S_{yx}^A$). Through subtracting the Normal Nernst signal $S_{yx}^N$, we can extract the anomalous Nernst signal $S_{yx}^A$, as described in Figure 3c. A large anomalous Nernst signal, $S_{yx}^A$ close to 0.15 $\mu V/K$ at 100 K is observed, much larger than those of conventional FM materials (the pure metal Fe, Co, Ni and single crystal Fe$_3$O$_4$)[33]. However, this value is comparable to that of topological antiferromagnet Mn$_3$Sn[34], but significantly smaller than those typical magnetic topological semimetals, such as Co$_3$Sn$_2$S$_2$[35], CoMnGa$_2$[36], and YbMnBi$_2$[37]. The anomalous $S_{yx}^A(T)$ at zero field is extracted by performing a linear extrapolation on the $S_{yx}^A(B)$ data, presented in Figure 3(e). $S_{yx}^A(T)$ increases below $T_C$, and then shows a broad peak near 100 K, similar to the observation in Co$_3$Sn$_2$S$_2$[26,35]. Upon entering the FM state, $S_{yx}^A$ exhibits a gradual increase as ferromagnetic order develops. This behavior reflects the enhancement of spin-polarized transport in the ordered phase. However, as temperature approaches absolute zero, the Mott relation [20,35] dictates that $S_{yx}^A$ vanishes linearly with temperature due to the diminishing entropy flow. Consequently, $S_{yx}^A$ first reaches a maximum at an intermediate temperature before decreasing monotonically towards zero at the lowest temperatures.

Based on the measured quantities $S_{yx}$, $\sigma_{yx}$, $\sigma_{xx}$ and $S_{xx}$, we can calculate the off-diagonal thermoelectric conductivity $\alpha_{yx}$ using the relation $\alpha_{yx} = \sigma_{xx}S_{yx} + \sigma_{yx}S_{xx}$, as displayed in Figure 3b. The field-dependent $\alpha_{yx}(B)$ also exhibits a pronounced hysteresis loop, manifesting the anomalous thermoelectrical components. After subtracting the linear term at high fields, we can derive the anomalous thermoelectric conductivity $\alpha_{yx}^A$ in Figure 3(d), which peaks at a maximum value of 0.075 A/m.K at 100 K, comparable to that observed in the vdW two-dimensional magnet Fe$_3$GeTe$_2$[20]. Figure 3f presents the temperature dependence of $\alpha_{yx}^A/T$ below Tc, revealing a characteristic $T\ln T$ scaling behavior above 60 K. This scaling, also reported in magnetic topological materials, including PrMn$_2$Ge$_2$[38] and Co$_{3-x}$Fe$_x$Sn$_2$S$_2$[39], suggests that the source of ANE in this system originates from the contribution of the Berry curvature.

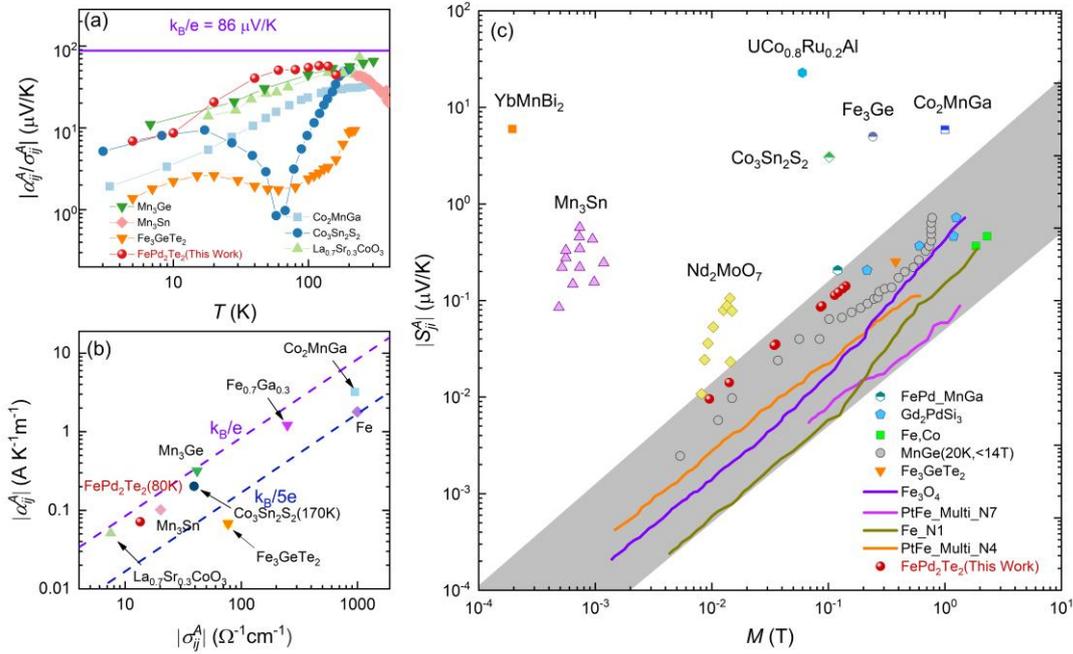

Figure 4 Comparison of different magnetic materials with FePd$_2$Te$_2$: (a) Temperature dependence of the ratio $|\alpha_{ij}^A/\sigma_{ij}^A|$; (b) The $|\alpha_{ij}^A|$ as a function of $|\sigma_{ij}^A|$ at room temperature, while Co$_3$Sn$_2$S$_2$ and FePd$_2$Te$_2$ are taken at 170 K and 80 K, respectively, due to the magnetic order. (c)The dependence of $S_{yx}^A$ below $T_C$ on the magnetization $M$

in FePd$_2$Te$_2$, conventional ferromagnets, such as FMs Fe, Co, and Fe$_3$O$_4$, and topological magnets, including Fe$_3$GeTe$_2$, Mn$_3$Sn, and Co$_3$Sn$_2$S$_2$, et.al. [40].

To further investigate the Berry curvature contribution to anomalous transverse transport in FePd$_2$Te$_2$, we systematically compared the $|\alpha_{ij}^A/\sigma_{ij}^A|$ ratio with various topological magnetic materials (Figure 4a-4b). Remarkably, the $|\alpha_{ij}^A/\sigma_{ij}^A|$ ratio in these systems can be limited by a saturation value of $k_B/e = 86\mu$V/K, providing further evidence that both the intrinsic anomalous Hall and Nernst effects originate from Berry curvature. Figure 4(b) demonstrates the intrinsic character of $\alpha_{ij}^A$ in FePd$_2$Te$_2$ through comparing with representative magnetic materials. The observed ratio $|\alpha_{ij}^A/\sigma_{ij}^A|$ falls between $k_B/e$ and $k_B/5e$--the nature units of the ratio of these correlated quantities, confirming the dominant role of Berry curvature in the ANE[36]. Note, while the $|S_{ij}| \propto \mu_0 M$ relation for FePd$_2$Te$_2$ lies at the boundary of this scaling regime (Figure 4c), its magnitude still exceeds those of conventional FM materials. This feature highlights the crucial influence of Berry curvature on the anomalous transverse thermoelectric effects.

## Conclusion

We have successfully synthesized high-quality single crystals FePd$_2$Te$_2$, a 2D-vdW magnet, and performed systematic investigations of its anomalous thermoelectric properties. Remarkably, below its $T_C$, FePd$_2$Te$_2$ exhibits robust zero-field AHE and ANE, accompanied by a pronounced hysteresis loop. In addition, the ratio $|\alpha_{ij}^A/\sigma_{ij}^A|$ approaches the fundamental limit of $k_B/e = 86\mu$V/K- the natural units for these transverse coefficients. These results indicate Berry curvature-dominated transverse transport mechanisms in this material. Our works establish this air-stable material as a unique platform for exploring low-dimensional magnetism and advancing next-generation thermoelectric spintronic applications.

**Acknowledgments**
This work was supported by the Hangzhou Joint Fund of the Zhejiang Provincial Natural Science Foundation of China (under Grants No. LHZSZ24A040001) and the Nature Science funding (NSF) of China (under Grants No. U1932155, 12274109).
# The authors contributed equally to this work.
Corresponding author: yklee@hznu.edu.cn
**Competing interests:** The authors declare that they have no competing interests.